# Velocity measurements of a dilute particulate suspension over and through a porous medium model

Eileen A. Haffner, Parisa Mirbod*

Department of Mechanical and Industrial Engineering, University of Illinois at Chicago,
842 W. Taylor Street, Chicago, IL, 60607, USA

**Abstract**

We experimentally examine pressure-driven flows of 1%, 3%, and 5% dilute suspensions over and through a porous media model. The flow of non-colloidal, non-Brownian suspensions of rigid and spherical particles suspended in a Newtonian fluid is considered at very low Reynolds numbers. The model of porous media consists of square arrays of rods oriented across the flow in a rectangular channel. Systematic experiments using high-spatial-resolution planar particle image velocimetry (PIV) and index-matching techniques are conducted to accurately measure the velocity measurements of both very dilute and solvent flows inside and on top of the porous media model. We found that for 1%, 3%, and 5% dilute suspensions the fully-developed velocity profile inside the free-flow region are well predicted by the exact solution derived from coupling the Navier-Stokes equation within the free flow-region and the volume-averaged Navier Stokes (VANS) equation for the porous media. We further analyze the velocity and shear rate at the suspension-porous interface and compare these data with those of pure suspending fluid and the related analytical solutions. The exact solution is used to define parameters necessary to calculate key values to analyze the porous media/fluid interaction such as Darcy velocity, penetration depth, and fractional ratios of the mass flow rate. These parameters are comparable between the solvent, dilute suspensions, and exact solution. However, we found clear effects between the solvent and the suspensions which shows different physical phenomenon occurring when particles are introduced into a flow moving over and through a porous media.

Key words; Very dilute suspensions, porous media model, volume-averaged Navier-Stokes

---

* Corresponding author.
 Email address: pmirbod@uic.edu (P. Mirbod).

# 1. Introduction

Suspension flows and their interaction over a porous surface are coupled between the properties of the suspension and the structure of the porous media. However, research into their behavior has bifurcated into two directions: suspension flows in geometries with smooth walls and flows of pure fluid over porous media. In this study, we examine the coupled flow of suspensions over and through a porous media model in a channel and analyze, in detail, the slip velocity, shear rate, and slip-coefficient at the interface between a porous media and a free flow region containing a dilute suspension flow.

Porous media has become prevalent in many fields spanning from biological findings to microfluidics and even environmental engineering systems. For example, flow over and through porous media has been examined in environmental phenomena such as flows over sediment beds [1], coral reefs and submerged vegetation canopies [2], and crop canopies and forests [3]. Flow over various biological porous media has been studied for different applications [4]. Also, porous media in microfluidics such as flow over carbon nanotube (CNT) forests [5], and flow past polymer brushes [6, 7] have been thoroughly researched. Some studies were reported using laser Doppler anemometry (LDA) to closely examine the flow behavior at the boundary of the pure fluid region and the porous media [8, 9]. Using LDA and manipulating the Darcy-Brinkman equation, it was found that the shear rate can be expressed as

$$\left.\frac{du}{dy}\right|_{y=0} = \frac{\alpha}{\sqrt{K\left(\frac{\mu_e}{\mu}\right)}} (u_s - u_D) \qquad (1.1)$$

where α is the slip-coefficient, $u_D$ is the Darcy or the superficial velocity, and $u_s$ is the slip velocity. They also compared eq. (1.1) with the shear rate proposed by Beavers and Joseph (1967) [10], using Darcy equation, and showed that the value of slip-coefficient α can be defined as $\alpha = (\mu/\mu_e)^{1/2}$. It has also been observed that α depends on parameters including the interface location, surface microstructure, porous media porosity ε, Reynolds number based on the Darcy velocity $u_D = -K\nabla p/\mu$, channel gap, and the bulk flow direction. James and Davis (2001) later introduced a dimensionless interfacial slip velocity, $u_s/(\dot{\gamma}\sqrt{K})$, where $\dot{\gamma} = du/dy|_{y=0}$ is the shear rate at the fluid-porous interface. This dimensionless quantity has been examined in several literatures relating to Newtonian fluid over porous media [11, 12]. Recently, using both theoretical methods and direct numerical simulations (DNSs), the impact of porosity on the pure fluid flow over porous media has been studied [54].

From a theoretical point of view the mathematical model is broken into two different regions: 1) the free-flow and 2) the porous media regions. The equation for the first region is the two-dimensional (2D) Navier-Stoke equation for fully developed, incompressible flow. There are two models used to describe the flow through the second region. The first is using Darcy's law, $Q = -K\nabla p/\mu$, where $Q$ is the volume flow rate through the porous media, $K$ is the permeability of the porous media, $\nabla p$ is the pressure drop through the media, and $\mu$ is the viscosity of the fluid. The other model combines Darcy's law and the Stokes equation [13]. For both approaches the difficulty then lies in prescribing the boundary conditions at the interface between the two regions. To maintain continuity the velocity and the shear stress has to be the same at the interface [12]. Equation 1.1 shows the microscopic description where properties must be calculated at each point, whereas in a mesoscopic approach one uses the local volume-averaged equations to describe the flow over an entire region [14]. More details regarding various theoretical models used to examine the flow over and through porous media and their comparison can be found in Bottaro (2019) [14]. The limitation of these models is that they only describe pure fluid and its interaction with porous media.

In addition to theoretical simulations and the experimental technique applied by Beavers and Joseph, LDA; laser doppler velocimetry (LDV) has also been used to study the flow dynamics through porous media. These known optical techniques, typically used in concurrence with refraction index matching (RIM) where the working fluid has the same index of refraction of the porous media so the study of the flow within the porous media would be visible. LDV and LDA both have been employed to study flow through packed beads at low porous Reynolds numbers to measure the porous velocity and study the development of inertial and drag within the pores [15, 16]. LDV has also been used to study higher Reynolds number flows as they exit a porous foam [17]. Their results provide velocity information in a small region which is ideal for examining the flow within the pores but does not allow the measurements across the entire porous media.

More recently another optical experimental method, particle image velocimetry (PIV) in conjunction with RIM techniques, have been used extensively to define the velocity distribution on top of and within porous media. This technique has been found to be more useful in the application of studying porous media since it has the capability of measuring two components of velocity which allows for more detailed data. While LDV is a point measurement that only provides local measurements in one direction [17]. For LDA experiments, knowledge of the porous media structure must be known which makes it obsolete for random porous media [18]. However, PIV has the ability to measure far field as well as local velocity profiles it has become the more dominate experimental technique to study flow over and through porous media.

PIV has been used in many applications to analyze various properties of the flow within the pores of the media, as well as porous media bounded by different free flow regions. PIV experiments, similar to the LDV and LDA studies, have examined viscous flow through packed beads [19-22]. Some efforts have been made to validate the volume averaging method within the pores of the beads [19,20]. Furthermore, it was found that the velocity within these pores was dependent on the geometry and configuration of the packed beads [22]. Along with packed beads, PIV has also been used to study the properties of the interface between a free flow region and porous media. The slip velocity, which is defined as the velocity at the fluid-porous interface, was determined for both Couttte flows [6, 23] and pressure-driven flows [24, 25]. Tachie et al. (2003) [23] compared the slip velocity obtained through PIV experiments to a model developed by Brinkman (1949) and found that the permeability along with the transverse velocity needed to be considered in the model to better simulate the flow. Other studies confirm this concept that the slip velocity is dependent on the permeability [25]. It has also been found that both the slip velocity and the free-flow velocity are dependent on the thickness of the porous media [24]. All of these studies are for pure Newtonian fluid over porous media, to the best of the author's knowledge there are no studies on examining very dilute suspension flows over and through a porous media model. The aim of this study is to characterize the impact of adding a small number of particles in the flow and examine this impact on the velocity of flow over and through a porous structure.

Suspension flows are encountered in phenomena such as oil, blood flow and pharmaceutical industries [26-28]. Several experimental techniques have been developed for the flow of rigid, spherical particles in confined, smooth geometries. For example, laser-Doppler anemometry (LDA) [29, 30] and laser-Doppler velocimetry (LDV) [31-33] were used to study suspension flows. Nuclear magnetic resonance imaging (NMR) has also been employed to study the suspension flows in various geometries [34-38]. Moreover, PIV has be utilized as a technique to analyze suspension flows within a rectangular channel containing smooth boundaries [39-42]. One of the main observations for higher concentrated suspensions ($> 20\%$) is that the particles move to the path of least resistance, meaning they migrate to the center of the flow field. This creates a local increase in concentration and viscosity of the flow in that region, which has been observed in both pressure-driven and shear flows. This increase causes a reduction in velocity which results in a blunted velocity profile in these flows. To the author's knowledge, there is a lack of in-depth experimental analysis on flow of very dilute suspensions over a porous structure.

From a theoretical/modeling perspective, there are two different continuum models to describe the dynamics of suspension flows. The first model, the diffusive flux model (DFM), introduced by Leighton and Acrivos (1987) [43] and Phillips et al. (1997) [44], is phenomenological and involves the diffusion fluxes of particles due to particle collisions and the spatial variation in the viscosity. The second model, the suspension balance model (SBM) is rooted in the conservation equations of mass, momentum, and energy for the fluid and particle phases [43-45]. The normal stress differences to handle curvilinear flows were considered in a research paper by Morris and Boulay (1999) [46]. This model has since been revisited by Nott et al. (2011) [47]. These models have mostly succeeded in predicting qualitative features of the migration process and quantitative steady-state velocity and concentration profiles in different geometries. For instance, the velocity and concentration profiles for 2D steady-state of both Brownian and non-Brownian suspension flow in a channel with smooth walls have been examined both theoretically and experimentally by [43, 46, 48-50]. Migration of rigid, spherical particles has been investigated using different methods including Stokesian dynamics (SD), force coupling method (FCM) and immersed boundary method (IBM). A comprehensive review of various methods has been reported recently in an annual review by Maxey (2017) [51].

In this work, we specifically examine dilute suspension flows and their behavior over and through a porous media model. In particular, very dilute suspensions with the bulk volume fraction $\phi_b \approx$ 1%, 3%, and 5% are examined by high-spatial-resolution planar PIV measurements. The 2D velocity in a plane is measured which allows for the calculation of the slip velocity, shear rate, and slip-coefficients. The experimental data then allowed us to provide useful data on the phenomenological constants used in the existing models for pure fluid over porous media and explore their validity for very dilute suspensions. We also compared the experimental results of both pure fluid and dilute suspensions with the exact solution for flow over porous media determined in our previous work [52]. The experimental set-up is described in section 2, the results are presented and discussed in section 3, and concluding remarks are presented in section 4.

## 2. Experimental Procedure
### *2.1. Neutrally buoyant suspension*

The particles used for the suspension fluid are polymethyl methacrylate (PMMA), monodispersed particles with a mean dimeter of 82 μm (75 μm≤ $d_p$ ≤ 90 μm) and a density of 1200 kg/m³ (Cospheric, LLC). This yields a gap to particle ratio of $L/d_p = 13.4$ in the free flow region. A density- and refractive-index-matched solvent was prepared by slightly modifying a recipe proposed by Lyon and Leal (1998a) [32]. The solvent consists of 55 wt% Triton X-100 (Sigma-Aldrich), 25 wt% 1,6-dibromohexane (Sigma-Aldrich), 10 wt% UCON oil 75-H-450 (Dow Chemical Company), and 10 wt% UCON oil 75-H-90000 (Dow Chemical Company). The Lyon and Leal (1998a) solution had to be modified because the PMMA particles used in the current study is higher than those used in Lyon and Leal's paper, and the ambient temperature in the lab where the experiments were conducted was T=23.4 ± 0.2°C for this study which is higher than the temperature reported in Lyon and Leal's paper. To determine the new mixture, the density was measured for each fluid using a hydrometer and the index of refraction was measured using a refractometer (Cole-Parmer), which were both measured at the ambient lab temperature. The final solution's density and refractive index were measured to get the necessary concentration that would work for the PMMA particles we had and the ambient temperature we had for these experiments. The resulting index of refraction of the solvent is, $n = 1.4897$, and its density is 1190 kg/m³. The solvent viscosity, $\mu_f = 0.288$ Pa.s at $T = 22°C$ that was measured with a concentric cylinder rheometer (Discovery Hybrid Rheometer-2 (DHR-2), TA Instruments). The suspension is prepared at various values of particle volume fraction obtained as

$$\phi_b = \frac{\pi d_p^3 N}{6 V_{total}} \quad (2.1)$$

where $N$ is the total number of particles and $V_{total}$ is the volume of the flow channel.

There is a slight mismatch between the density of the PMMA particles and the solvent. To ensure the particle remain suspended during testing the hindered settling was calculated. This settling would mathematically consider the mismatch of the fluid densities along with the interaction of the suspension particles as higher particles would travel down to the bottom wall [53]. It was found that the settling velocity of the PMMA particles was less than 0.04% of the average velocity within the channel, allowing this solution to be stable long enough for the tests performed.

In this study, we were particularly interested in flows with very low Reynolds number, allowing the suspension to be in the creeping flow regime. Due to the complexity of the flow there are three different Reynolds numbers used to characterize this flow. The first Reynolds number is based on the PMMA particle size, $Re_p = \frac{4\rho_f a^3}{3\mu L^2}|U_{max}|$, where $\rho_f$ is the density of the fluid, $\mu$ is the viscosity of the Newtonian fluid, $L$ is half the height of the free flow region, $a$ is the average radius of the PMMA particles, and $U_{max}$ is the streamwise maximum velocity within the flow channel [32]. For this study, $Re_p \sim O(10^{-6})$, therefore the inertia of the particles can be neglected. The second Reynolds number to consider is the Reynolds number based on the porous media. The equation for this is $Re_L = \frac{\rho_f d U_b}{\mu(\phi)}$, where $d$ is the diameter of the rods making up the porous media, and $U_b$ is the streamwise bulk velocity $\left(U_b = \frac{1}{H_T}\int_{-H}^{2L} u(y)dy\right)$, and $\mu(\phi)$ is the Krieger viscosity. Krieger proposed a concentration-based viscosity derived from parameter fitting of rheological experiments. The resulting relationship for this viscosity is,

$$\mu(\phi) = \mu\left(1 - \frac{\phi_b}{\phi_m}\right)^{-\beta} \qquad (2.2)$$

where $\mu$ is the viscosity of the pure fluid, $\phi_b$ is the bulk particle volume fraction of the initial suspension, $\phi_m$ is the maximum packing particle volume fraction, and $\beta$ is a rheological fitting parameter [55]. For our case we consider $\phi_m = 0.68$ and $\beta = 2$ [50], which results in porous Reynolds number of $Re_L \sim O(10^{-1})$. The final Reynolds number uses half the free flow region as the characteristic length, the resulting equation for the suspension Reynolds number is $Re_S = \frac{\rho_f L U_b}{\mu(\phi)}$. This Reynolds number ranged from 0.029 to 0.035. It should be noted that the Stokes number, defined as $St = m_p\dot{\gamma}/3\pi\mu d_p$ where $m_p$ is the mass of the particles and $\dot{\gamma} = U_{max}/L$ is the shear rate [50, 56]. For the current study it was calculated that, $St \sim O(10^{-5})$ meaning that the motion of the suspension particles is solely governed by the velocity of the fluid flow. The final dimensionless parameter to consider is the Péclet number, $Pe = 3\pi\mu d^3\dot{\gamma}/4k_B T$. For our experiments it is $O(10^9)$ for the range of shear rate used in this work. These represent that the particles are non-Brownian ($Pe \gg 1$), the inertia of particles is negligible $Re_p \ll 1$, and the particle responses quickly to the change in the flow and behaves as a tracer (i.e., St≪1).

TABLE I. Experimental parameters for the various suspension fluids through and above a porous media model.

| $\phi$ | $U_b$ | $Re_P$ | $Re_S$ | $Re_L$ | $\varepsilon$ | $H$ (cm) | $l$ (cm) | $d$ (cm) | $H_T$ (cm) | $L_T$ (cm) | $W$ (cm) |
|---|---|---|---|---|---|---|---|---|---|---|---|
| 0 | 0.789 | 4.48e-6 | 0.035 | 0.024 | | | | | | | |
| 0.01 | 0.786 | 4.22e-6 | 0.034 | 0.023 | 0.9 | 0.5 | 28.4 | 0.15 | 0.7 | 100 | 2.5 |
| 0.03 | 0.722 | 4.00e-6 | 0.030 | 0.020 | | | | | | | |
| 0.05 | 0.740 | 3.93e-6 | 0.029 | 0.020 | | | | | | | |

(a)

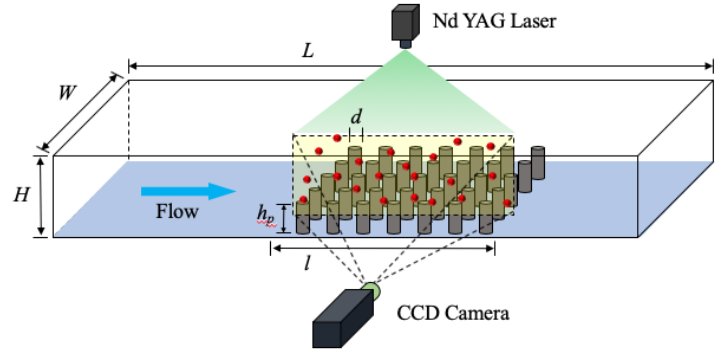

(b)          (c)

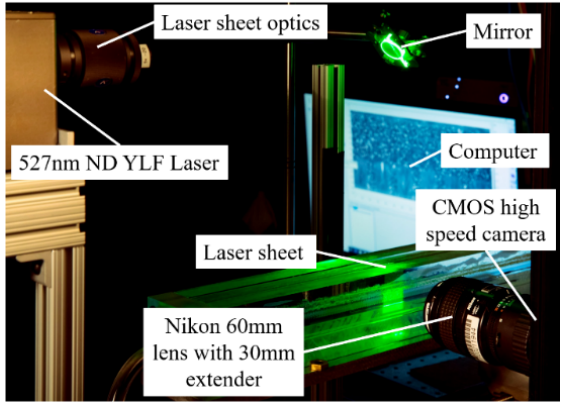
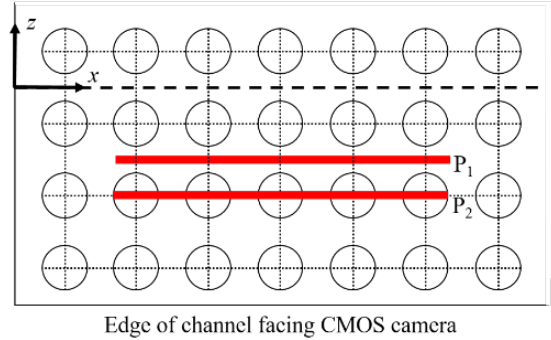

FIGURE 1. (a, b) A schematic of the flow channel and the PIV experimental set up, and (c) schematic of the top view of the porous media. $P_1$ corresponds to the laser plane within the porous media, and $P_2$ shows the laser plane on top of the porous media.

Table I reports the typical parameters used in this study: the average magnitude of the bulk velocity, $U_b$, of each experiment, the height of the porous media model, $H$, the length of the porous media model, $l$, the diameter of the rods, $d$, and all of the various Reynolds number required for this study: $Re_P$, $Re_S$ and $Re_L$. The resulting velocity information is the area-averaged from the sides of the rods in the porous model. The magnitude of the velocity was expressed as $U(y) = \sqrt{u(y)^2 + v(y)^2}$ and the average bulk velocity is defined by, $U_b = \frac{1}{H_T}\int_{-H}^{2L} U(y)dy$.

### *2.2. Experimental setup, PIV system, and measurements procedure*

The flow channel for this test was constructed from high quality, clear, cast acrylic as can be seen in Fig. 1(a) and 1(b) (ACME Plastics, Inc). The porous media model used was constructed via a 3D printer (ProJet, Laser Concepts) using VisiJet Crystal. The porous media model consists of rods in a square array aligned perpendicular to the flow direction. The rods have a diameter of 0.15 cm and the space between the sides of the rods is 0.27 cm. The porosity of this model is 90% in which a porosity of 100% would be equivalent to an empty (smooth) channel. The permeability was calculated using a series of equations proposed by Tamayol and Bahrami (2011) [57] for fibers in a square array. The resulting permeability, $K$, based on the rod diameter and the porosity is 7.93x10$^{-7}$m$^2$ [57]. The flow loop uses a peristaltic pump (MasterFlex L/S, Cole-Parmer) with two pump heads (Easy Load II, Cole-Parmer). The pump heads feed into a Y fitting which connects to a dampener (Masterflex Pulse dampener, Cole-Parmer). The combination of the Y fitting and the dampener helps reduce any pulsating from the pump, so the resulting flow is smooth and at a constant speed when it enters the flow channel. A flowmeter (mini CORI-FLOW, Bronkhorst USA) is placed in between the end of the dampener and the start of the flow channel. This allows for a real time measurement of the fluid flow rate, density, and temperature. Measuring the flow rate during experiments can act as a validation method to ensure the velocity vectors from the PIV corresponds to the same flow rate that was measured. A reservoir fitted with two outlets and one inlet was set between the outlet of the flow channel and before the peristaltic pump which closes the flow loop. For each experiment, the suspension fluid was mixed for two hours on a magnetic mixer within the reservoir. Then the fluid is put into a vacuum chamber and subjected to a pressure of -30 inHg, and left for an addition hour to remove any ambient air bubbles from the suspension. Once the pump was started, the flow was left to stabilize through the flow loop for 30 minutes to reach a steady-state condition before any data was collected.

Flow-field measurements were developed using a LaVision Flow Master particle image velocimetry (PIV) system. This consisted of a 527 nm Nd-YLF (Photonics Industries, DM20-527) to illuminate the tracer particles which are seeded within the flow, as well as a high-speed complementary metal-oxide semiconductor (CMOS) camera with a  pixel field of view and a full-scale resolution. A band-pass filter (BP532 10) was placed over a 60-mm-focal-length lens attached to the camera in order to minimize spurious reflections and enhance the particle signal-to-noise ratio. To be able to better resolve the velocity vectors between the rods of the porous media a 36 mm extension tube was placed between the CMOS camera and the 60 mm lens. All hardware timing and software were handled using the DaVis10.0.5 software package on the Intel(R) Xeon(R) CPU with 32.0 GB RAM and 64-bit operating system. Fig. 1(b) reveals the experimental setup.

## *2.3. Velocity measurements using PIV protocol*

Fluorescent seeding particles with an average diameter of 2μm and a density of 1300 kg/m$^3$ were used as tracer particles for PIV measurements (Cospheric, LLC). The particles peak excitation wavelength was 537 nm and their peak emission wavelength was 594nm. To minimize error associated with the PIV, the tracer particles remain small enough to ensure they stay along the natural streamlines of the fluid, but not so small, that their motion subjected to Brownian affects. These tracer particles are at a very low concentration (lower than that of the suspension particles) so that they will not be interacting with one another. The Stokes' drag law was employed to find the settling velocity of a single tracer particle which was determined to be less than 0.0008% of the average velocity. Therefore, the tracer particles are stable within the fluid. The Péclet number for seeding particles was calculated to be, $Pe_{seed} = 3\pi\mu d_{seed}^3 \dot{\gamma}/4k_B T = O(10^5)$. Therefore, the Brownian motion of seeding particles can also be neglected [58].

Following this section, we discuss the necessary sample size to take the PIV data, the location to place the porous media model, fully-development of both solvent and dilute suspensions in the channel over the porous model. In the current study, the data were collected in two different planes, which are illustrated in Fig. 1(c). The use of two data collection planes averaged together would be enough to fully resolve the flow through and over the porous media which was discussed through Arthur et al. (2009) [25]. Based on these finding, the plane within the porous media is labeled P$_1$ and the plane on top of the porous media is labeled P$_2$. Velocity data was processed for each of these planes and then averaged together to get the final velocity profile that described the flow throughout the porous media (see [25] for more details).

Before conducting experiments with a porous media model, we examined the suspension flow in a smooth channel to characterize the exact location where the porous slides should be placed so that the flow is fully developed before encountering the porous media. To determine the location of the porous media from the inlet of the flow channel, we performed a test using a suspension with a bulk concentration of $\phi_b = 0.03$ in a smooth channel (no porous media present) to ensure the flow was fully developed before meeting the start of the porous media. This test was performed at a flow rate of 69.1 mL/min. Fig. 2(a) shows velocity profiles gathered using PIV measurements at various axial locations downstream from the inlet. As can be seen in the Fig. 2(a), these velocity profiles from the data agree well with the velocity profile obtained using the exact solution for flow within parallel plates as derived from the steady, incompressible continuity and Navier-Stokes equations. This indicates the flow at all axial locations were fully-developed. Therefore, we located the porous media 8 cm downstream from the inlet, which allowed for fully developed flow while it started to interact with the porous media.

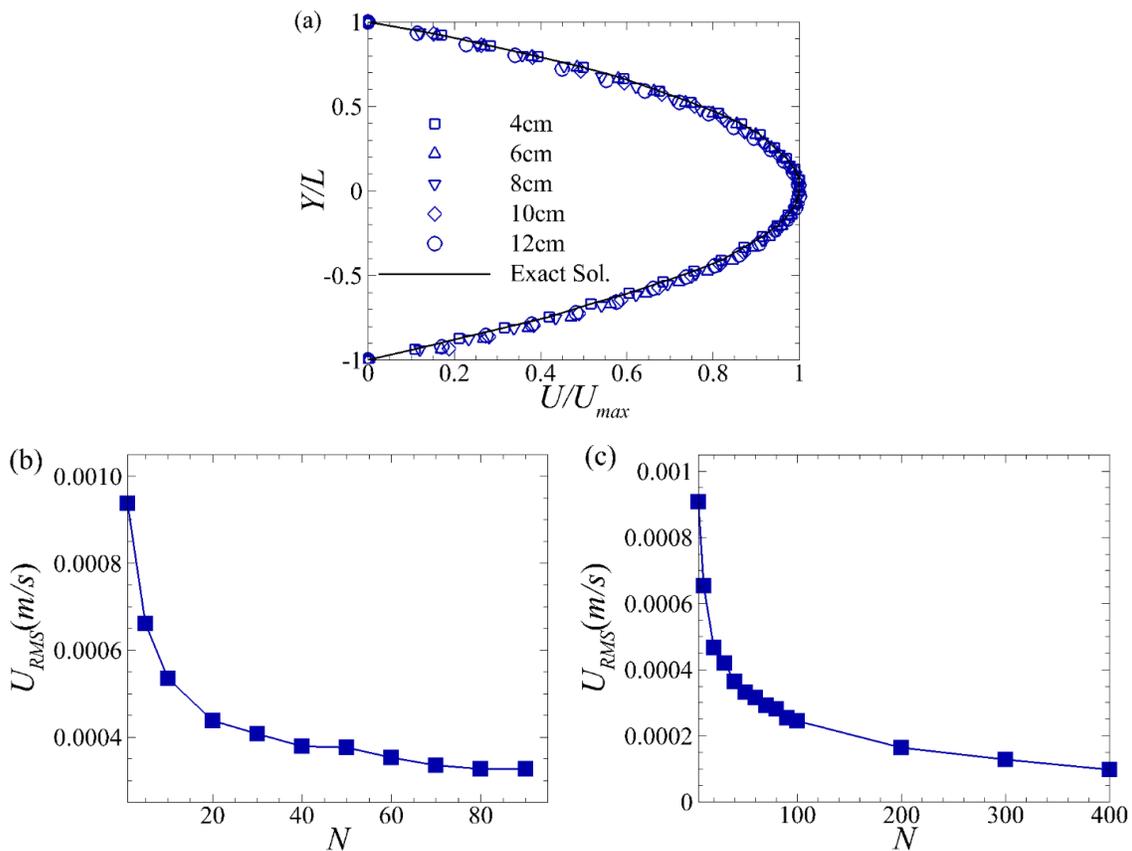

FIGURE 2. (a) Velocity profiles obtained through PIV experiments at different axial locations downstream of the flow channel inlet. The solid black line shows the exact solution for flow between parallel plates. Statistical convergence test using 3% suspension for (b) smooth channel, and (c) over the porous medial model.

Moreover, to evaluate the necessary sample size required for the statistical convergence of the PIV data, for both the smooth channel and the channel containing the porous media, a test with varying number of image pairs for a 3% suspension was performed. For the smooth channel, a test with 100 image pairs was taken at the inlet. The inlet was considered to be the "worst case scenario" for the PIV data, because that location has more mixing than further downstream. The data was then processed and averaged for varying amounts of image pairs up to the maximum 100 image pairs taken. The averaging of the full data set was considered the most accurate velocity vectors, therefore, the other averaged velocities for fewer image pairs were then subtracted from the averaged full data set. Fig. 2(b) shows the resulting root mean square (RMS) for the remaining velocity values from the difference between the averages using fewer image pairs to that of the full data set. This RMS shows how much the averaged smaller data sets vary from the full averaged data set. This graph trails off around 80 image pairs; meaning that if more image pairs were processed and averaged together, they would result with similar velocity vectors. This represents the required sample size necessary to obtain statistical convergence of the PIV data. A similar approach was conducted for the 3% suspension flow over the porous media. For this sample test, 500 image pairs were collected at the beginning of the porous media for the $P_1$ plane. Again, this is to simulate a "worst case scenario" so that data taken at any other place along the porous media would result in more clear data. Similarly, as shown in Fig. 2(c), we found that 300 image pairs are sufficient to get statistical convergence when the porous media model is located inside the channel.

The raw PIV data was taken at a frequency of 250 Hz with a laser input current of 20 amps, which resulted in an image with the highest signal to noise ratio. A typical raw PIV image can be seen in Fig. 3(a). The physical field of view (FOV) was approximately 21 mm by 14 mm. The PIV data is post-processed using a multi-pass approach. This started with a one-pass of a square interrogation window size of 64 x 64 pixels$^2$ with 50% overlap and concluded with four passes of a circular interrogation window size of 24 x 24 pixels$^2$ with 75% overlap. This resulted in a velocity vector every 0.07 mm. Fig. 3(b) shows the resulting streamwise velocity ($u$) contours for the solvent in the plane within the rods (i.e., $P_1$) and Fig. 3(c) shows the data collection in the plane on top of the rods (i.e., $P_2$). As can be seen in Fig. 3(b), the maximum average velocity occurs in the free-flow region, which is also observed in Fig. 3(c) for the $P_2$ plane. For both data collection planes, the maximum velocity appears on the top of the rods which decays as the flow moves through the rods. The velocity contours for 5% concentration suspension are shown in Fig 3(d) for P$_1$ and Fig. 3(e) for P$_2$. The contours resemble that of the solvent test; however, for both data collection planes the velocity distribution within the porous media is slightly different, showing jagged transitions between velocity contours. This could be due to visibility issues at the higher concentration of suspensions. It should be noted that the relative uncertainty of the velocity obtained from the PIV technique was determined to be at most 0.53% of the maximum velocity for both data on the plane within the porous media and for the data taken on top of the rods. These uncertainties were calculated based on the spacing between the velocity vectors as well as the standard deviation of the velocity data reported [59, 60].

The spanwise velocity contours ($v$) in the plane within and on top of the rods for both solvent and 5% suspension can be seen in Fig. 3(f-i). As can be expected, there is little to no transverse velocity within the rods for both cases (see Fig. 3(f),(h)). In the plane on top of the rods for the solvent fluid, Fig. 3(f), and 5% suspension, Fig. 3(h), the velocity has a peak at the corner of the rods that face the oncoming flow at the interface. Conversely, the peak velocity is negatively mirrored on the downstream corner. This effect is more prominent for the 5% suspension case.

To ensure the flow over and through the porous media model is fully developed, meaning the flow is periodic, we characterized our measured velocity data slightly below the interface between the porous media and the free flow region for both data collection planes. To visualize this, the velocity magnitude is plotted in a x-y plane at a location of $Y/L = -0.63$. This location was selected because it is one row of velocity vectors below the interface location. Fig. 4(a) shows the specific FOV windows where the data has been taken along the channel over the porous media. Fig. 4(b-d) shows the velocity data in the x-y plane inside the channel for both the P$_1$ and P$_2$ data collection planes from $x = 8cm$ to $x = 36.5$cm from the inlet of the flow channel. For the solvent (Fig. 4(b)), 1% suspension (Fig. 4(c)) and 3% suspension (Fig. 4(d)) the flow becomes periodic at $26cm < x < 34cm$ from the inlet of the channel for both data planes. Accordingly, we analyzed all of our data at $0.63 \leq x/l \leq 0.94$ to ensure all flows are fully developed and there is no change in the velocity magnitude along the x direction in that region. The data obtained in these FOV's is then averaged together to create all 1D velocity profiles discussed in the following section.

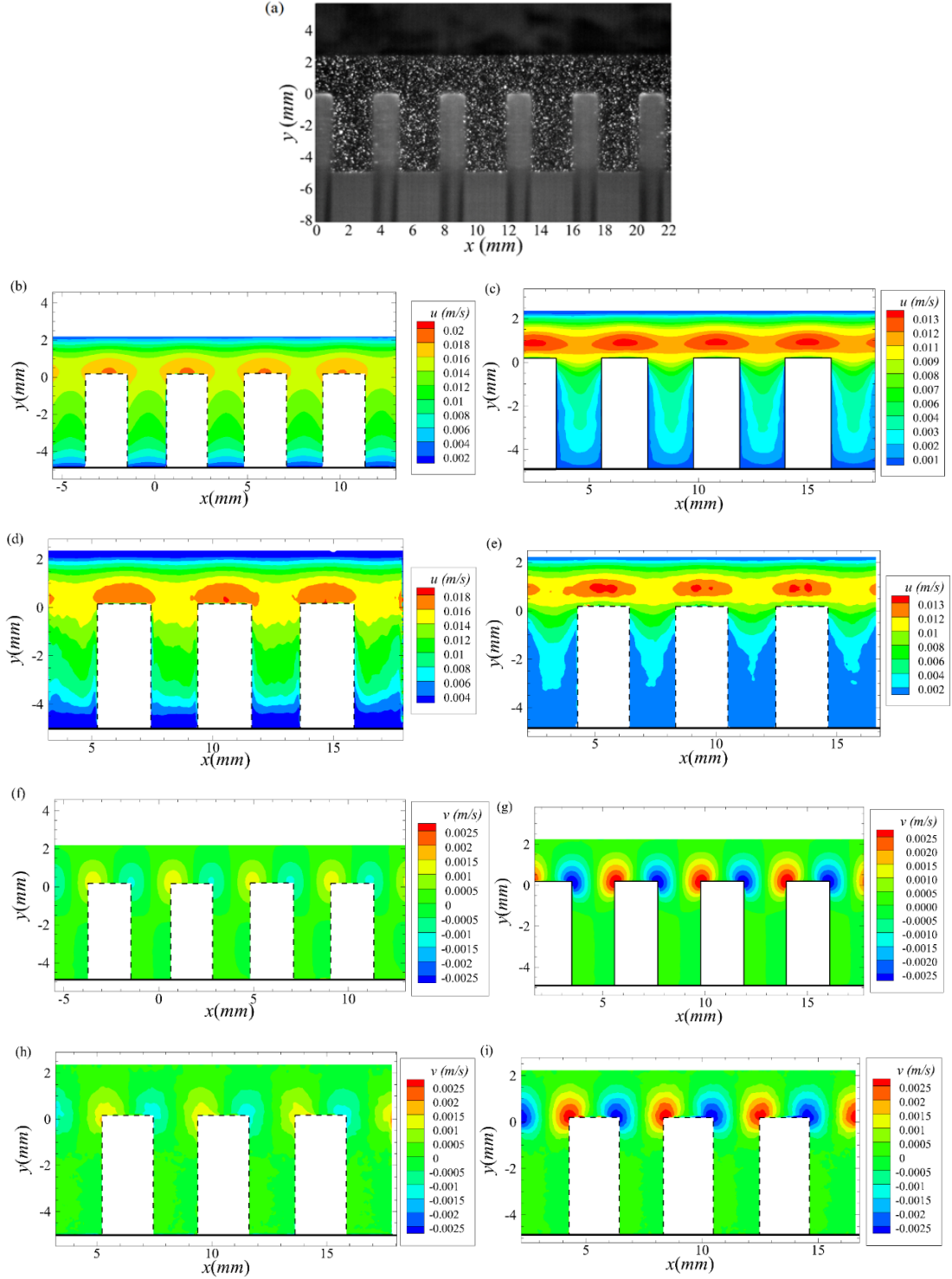

FIGURE 3. (a) Raw PIV image for a suspension with $\phi_b = 0.01$ for the $P_1$ plane. Velocity contours for the solvent for the (b) plane within the rods, $P_1$, and (c) the velocity contours for the plane on top of the rods, $P_2$. The velocity contours for the same data collections planes, (d) $P_1$ and (e) $P_2$ for $\phi_b = 0.05$. The velocity contours for the transverse velocity for the pure suspending fluid for the (f) $P_1$ and (g) $P_2$ planes as well as for the 5% suspension for the (h) $P_1$ and (i) $P_2$ planes.

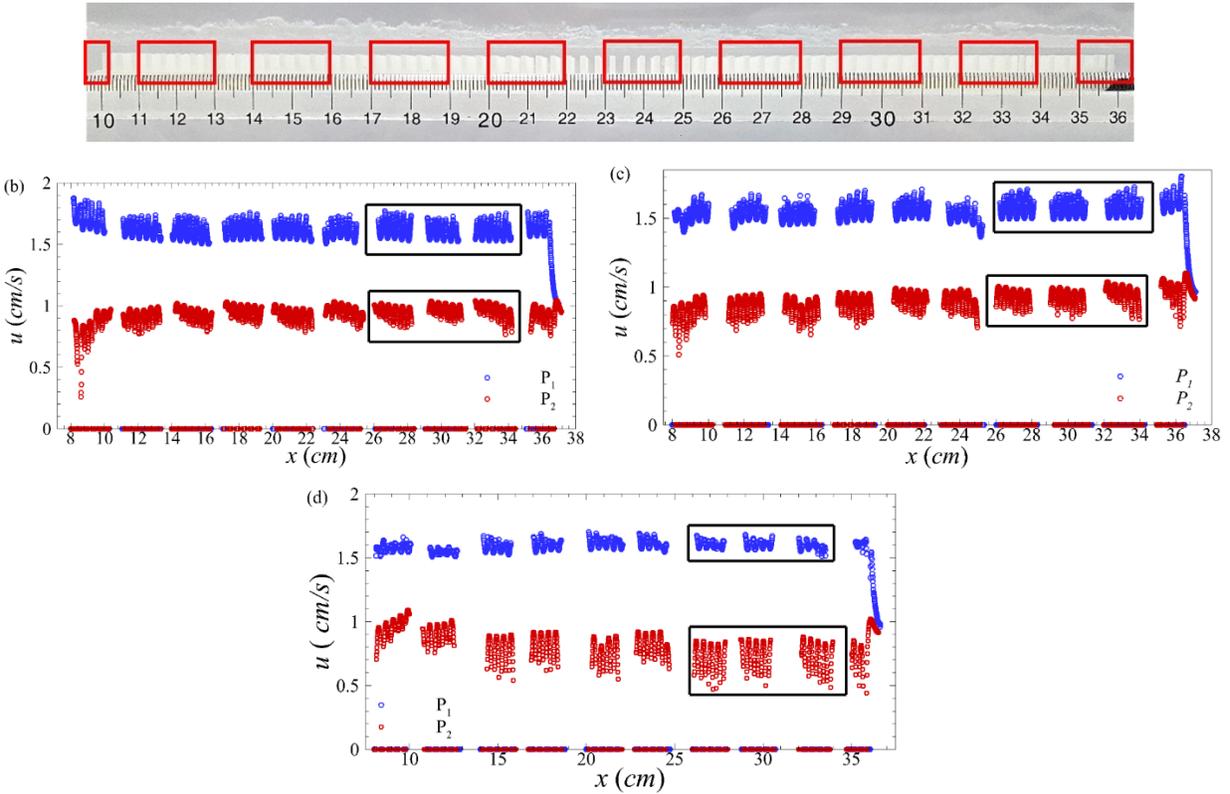

FIGURE 4. (a) Side view of channel containing porous media, the boxes are indicating location of windows where data was collected. To check the periodicity of the flow, (b) shows the streamwise velocity over the entire porous media for both data collections planes, $P_1$ and $P_2$ or the solvent, (c) 1% suspension, and (d) 3% suspension.

Previous literature has shown that the normalized velocity profile for a Newtonian fluid over porous media at different Reynolds numbers will collapse to the same normalized velocity profile trend, indicating that inertial effects are negligible [25]. To ensure that this phenomenon occurred in our experiments, the solvent case and the three dilute suspensions (1%, 3%, and 5%) were examined at three Reynolds numbers. Since fundamental physical parameters of the setup, including the channel size and the porous media structure, remained constant between experiments, we had to change the flow rate from approximately 60 mL/min to 80 mL/min to achieve these different Reynolds numbers. It should be noted that, for the dilute suspensions the suspension Reynolds numbers, $Re_S$, was considered. Fig. 5 shows the resulting normalized velocity profiles for these experiments. For all cases there was a good agreement between the Reynolds numbers. For the solvent case, Fig. 5(a), and the 1% suspension, Fig. 5(b), the profiles are almost perfect overlays of one another. In the higher concentrated suspensions, 3% and 5%, there are more discrepancies between the profiles for the three suspension Reynolds numbers, especially in the porous media. We believe this effect was due to variation in the velocity profiles because of particle interaction in the porous media. In the rest of our analyses we considered $Re = 0.035$ for the suspending fluid, $Re_S = 0.034$ for the 1%, $Re_S = 0.030$ for the 3%, and $Re_S = 0.029$ for 5% suspensions.

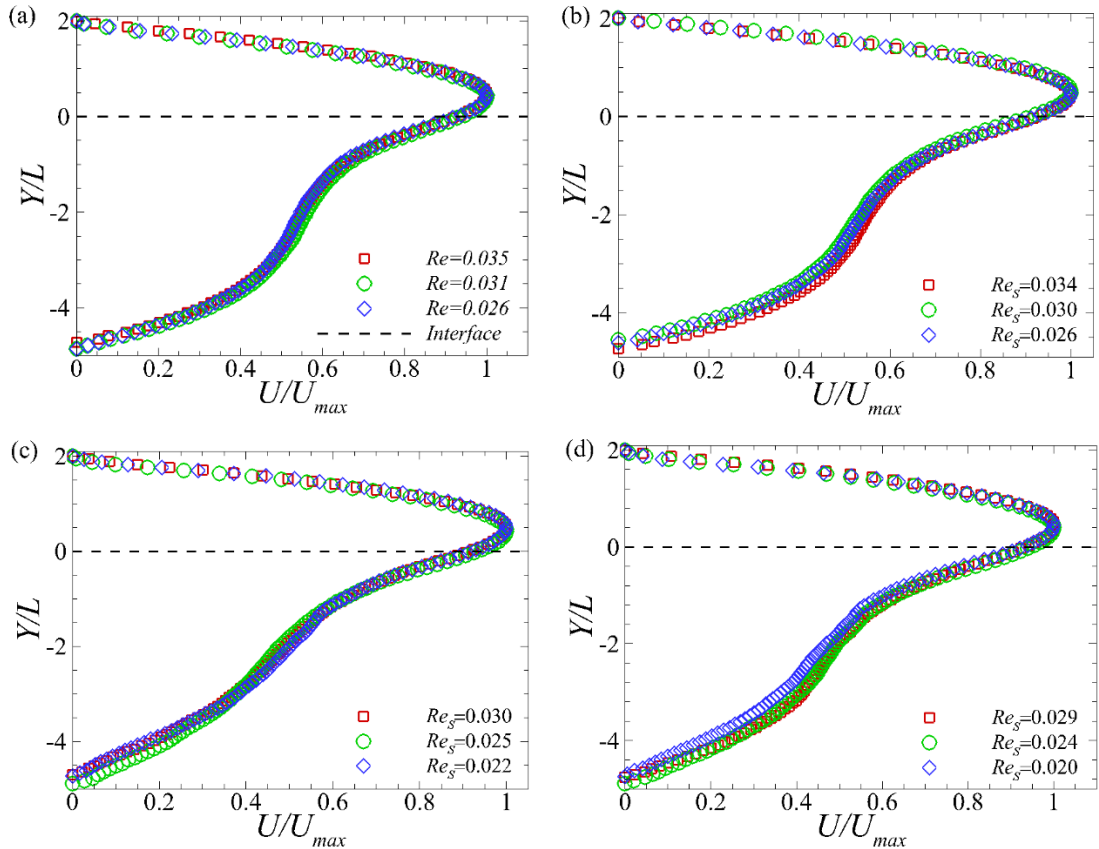

FIGURE 5: Effect of Reynolds numbers on flow through channel with the existence of the porous model for (a) solvent and for suspensions with (b) 1%, (c) 3%, and (d) 5% suspension. The Reynolds numbers are, 0.02-0.035, since they are low the flow is independent of inertia effects.

## 4. Results and discussion

In this study, we were not only interested in characterizing the very dilute suspensions' averaged velocity profiles describing the full flow field, but also their impact on each data collection planes (i.e., $P_1$ and $P_2$). Due to the 3D impact of the porous media, it was observed that there was a clear difference in velocity distribution between the data plane within the rods, $P_1$, and on top of the rods, $P_2$ (see Fig. 3(b-i)). This led to different properties at the fluid-porous interface which we examined in this article. We have compared both components of the velocity profile for the suspending fluid, and different initial bulk volume fractions of $\phi_b = 0.01, 0.03,$ and $0.05$ flowing over and through the porous model. Fig. 6(a-d) shows the streamwise, $u$, and spanwise, $v$, components of the velocity profiles for both the (a) $P_1$ and the (b) $P_2$ planes.

For all cases, it can be observed that the transvers velocity values ($v$ component of velocity) are much less than that of the streamwise velocity values ($u$ component of velocity). The transvers velocity is practically zero in all locations except for a small area around the interface between the porous media and the free flow region. We found that for all data collected in the $P_1$ plane, the maximum transverse velocity value was less than 10% of the maximum streamwise velocity value. However, in the $P_2$ plane the transverse velocity values were higher than in the $P_1$ plane, which led to $v$ component of velocities that were 12.5% to 14.5% of the $u$ component. This effect could be due to multiple of properties of either the porous media or the suspension fluid. Previous literature has reported that when the $v$ component of velocity is 10% less than the $u$ component, then it can be neglected. For our case in the $P_2$ plane this is not true, therefore, we must consider a velocity magnitude, $U(y)$, for the rest of the analyses.

Fig. 6(e) shows the velocity profile obtained by averaging of the $P_1$ and $P_2$ planes for the solvent fluid and the different diluted suspensions. As can be expected, the profile is parabolic in the free-flow region which then decays to the slip velocity, $U_s$, at the free flow-porous interface. All experimental cases show good agreement with one another within this free flow region. The maximum magnitude of velocity, $U_{max}$, occurs close to the interface. This is because of the resistance in the flow caused by the existence of the porous media [12]. As reported in our previous works for pure Newtonian fluid this effect depends on the thickness and the properties of the porous media [12, 61]. In general, we found that for dilute suspensions (i.e., $0.01 \leq \phi_b \leq 0.05$) passing over and through the porous media model, the flow has very similar flow physics to that of pure Newtonian fluid [62, 63], while it is slightly different for flow within the porous media. We also found that inside the porous media as the concentration increases the velocity is reduced. The exception to this is the 3% suspension test. This could be due a slight mismatch between the index matching of the solution to the particles. Thus, our experimental data reveals that, even for very dilute suspensions, the interactions of the rods with particles, the behavior of the particles within the rods, particle size, thickness, properties, and the 3D structure of the porous media model are all critical.

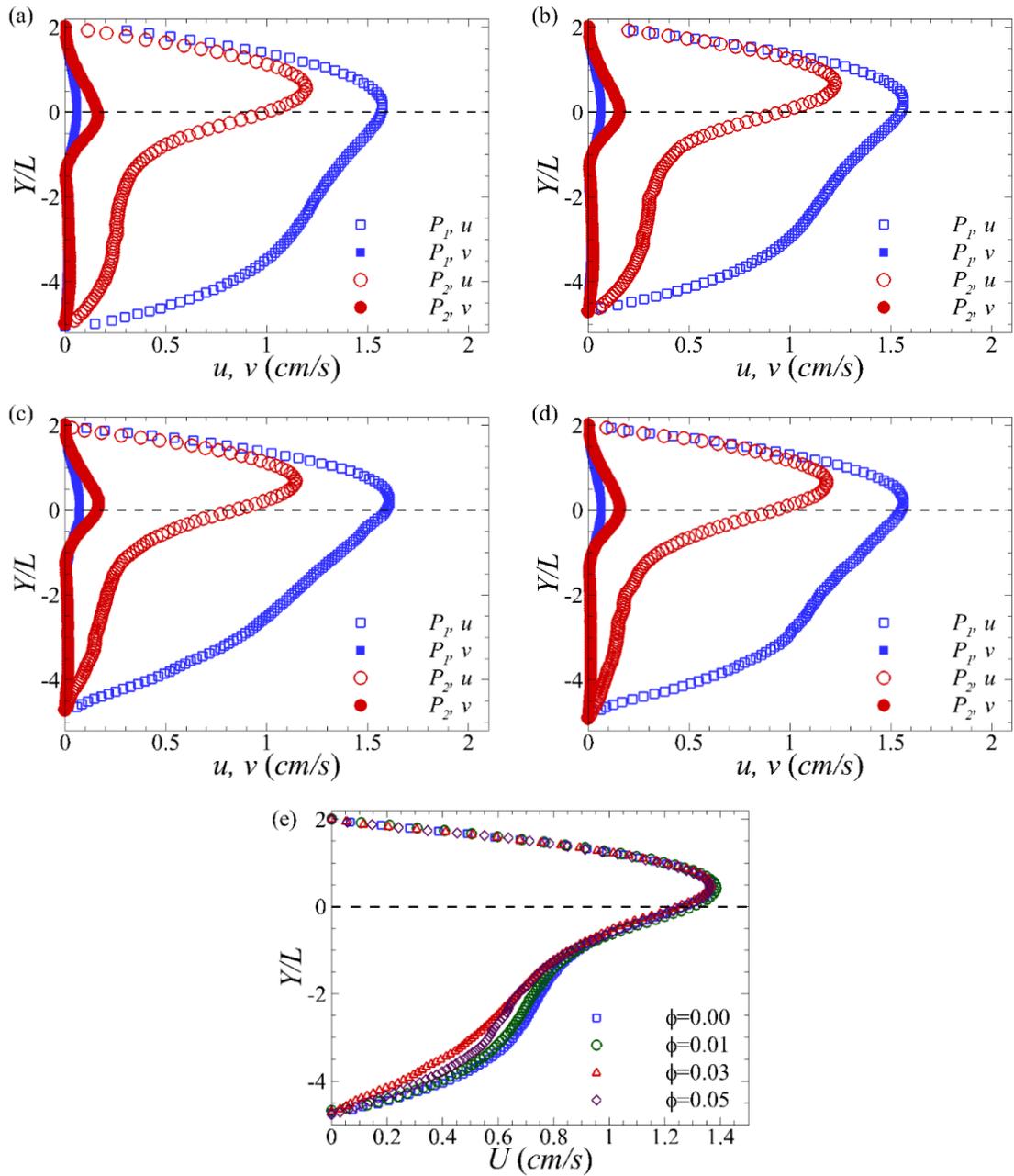

FIGURE 6. The spanwise, $v$, and streamwise, $u$, velocity profiles for (a) the plane within the rods and (b) the plane on top of the rods for pure suspending fluid and suspensions with different initial bulk volume fractions of $\phi_{bulk}$ = 1%, 3%, and 5%. (c) The normalized profiles for pure suspending fluid and various suspension of 0.01, 0.03 and 0.05.

In addition, we found that in all of our velocity analyses, the flow within the rods is not uniform. This effect is more evident when more particles are added to the flow. For the solvent, this is due to the wall effects that are similar to the data taken for pure Newtonian fluid moving over and through porous models reported by [24, 25, 64]. Moreover, as reported in these literatures for the thickness and the porosity of the porous media model considered in this work, the small boundary layers produced at the interface and on the lower surfaces of the channel might extend to result in non-uniform velocity profiles. These impacts have been proven to be vanished for pure Newtonian fluid motion over porous model by decreasing porosity of the porous media [12, 25].

Further, we examine the analytical results obtained from our previous theoretical/numerical work where an exact solution for pure Newtonian fluid through a channel were the bottom boundary had been replaced with porous media was derived [52]. The governing equation for the free-flow region is the unsteady, incompressible continuity and Navier-Stokes equations. These equations can be simplified through the assumptions that the flow is steady and one-dimensional. The volume-averaged Navier-Stokes (VANS) equations are then used to model the flow over and through the porous media [65, 66]. The resulting equations are nondimensionalized using the free-flow region half height, $L$, and $q = -(L^2/\mu)\, dp/dx$ as the characteristic length scale and velocity, respectively. The exact solution for the Newtonian fluid over a porous media boundary is then stated as,

$$\tilde{u}(\tilde{y}) = \frac{1}{\sigma^2} + C_1 e^{\sigma\sqrt{\varepsilon}\tilde{y}} + C_2 e^{-\sigma\sqrt{\varepsilon}\tilde{y}}, \qquad \tilde{y} \in [-\delta, 0] \tag{4.1}$$

$$\tilde{u}(\tilde{y}) = -\frac{\tilde{y}^2}{2} + \left(1 - \frac{\tilde{u}_s}{2}\right)\tilde{y} + \tilde{u}_s, \qquad \tilde{y} \in [0,2] \tag{4.2}$$

where

$$C_{1,2} = \pm \frac{1}{\sigma^2} \frac{(\sigma^2 \tilde{u}_s - 1)e^{\pm\sigma\sqrt{\varepsilon}\delta} + 1}{e^{\sigma\sqrt{\varepsilon}\delta} - e^{-\sigma\sqrt{\varepsilon}\delta}},$$

$$\tilde{u}_s = \frac{\sigma\sqrt{\varepsilon} - \sigma\sqrt{\varepsilon}\operatorname{sech}\sigma\sqrt{\varepsilon}\delta + \sigma^2 \tanh\sigma\sqrt{\varepsilon}\delta}{\left(1 + \frac{\tanh\sigma\sqrt{\varepsilon}\delta}{2\sigma\sqrt{\varepsilon}}\right)\sigma^3\sqrt{\varepsilon}} \tag{4.3}$$

Here $\tilde{u} = u/q$ is the dimensionless velocity, $\tilde{y} = y/L$ is the dimensionless y coordinate, $\tilde{u}_s = u_s/q$ is the dimensionless slip velocity, $\sigma = L/\sqrt{K}$ is the dimensionless permeability parameter, and $\delta = H/L$ is the dimensionless porous media thickness ratio [52].

Fig. 7 shows the normalized velocity profiles for pure fluid and the various concentrations along with the analytical profile from eq. (4.2) and (4.3). For the solvent, seen in Fig. 7(a), there is good alignment within the free-flow region and the top of the porous media. Within the porous media the experimental profile begins to deviate slightly from the theoretical profile. We believe this minor departure for the exact solution could be due to the fact that our porous media is not index matched to the suspending fluid, not allowing us to view velocity behind the rods. The experimental velocity for the 1% suspension, shown in Fig. 7(b), appears to be very similar to that of the solvent, but the velocity within the porous media is even further reduced causing a larger discrepancy from the exact solution. At the higher concentrations, 3% suspension shown in Fig. 7(c) and 5% suspension shown in Fig. 7(d), the variation of velocity within the porous media is much slower than the predicted profile from the exact solution. As mentioned before, we believe that this is due to the fact that the suspension particles are moving through the porous media leading to interactions between the particles and the rods which would reduce the velocity magnitudes in the region. The discrepancy between the model prediction and the experimental data is also because there were three-dimensional velocity affects in experiments that is not captured within the numerical analysis.

Different parameters were extracted from the magnitude velocity profiles plotted in the $P_1$ and $P_2$ planes as well as the averaged profile shown in Fig. 7. The maximum velocity was obtained from all these profiles, and was then normalized using the bulk velocity, $U_{max}/U_b$. As observed in Fig. 8(a), the normalized maximum velocity was relatively constant across for each concentration case tested. This was true for the $P_1$ and $P_2$ planes, as well as the full averaged velocity profile. We found that, on average, this normalized maximum velocity for the plane within the porous media, $P_1$, was 42.7% higher than the plane on top of the porous media, $P_2$. It should be noted this might be solely true for the porous media model used in this study; however, previous studies by showed that the values of $U_{max}/U_b$ strongly depended on the thickness, porosity and permeability of the porous media [12, 24]. Herein, we found for very dilute suspensions, the normalized maximum velocity does not depend on the suspension concentration. However, the impact of the porous thickness and properties need to be further investigated.

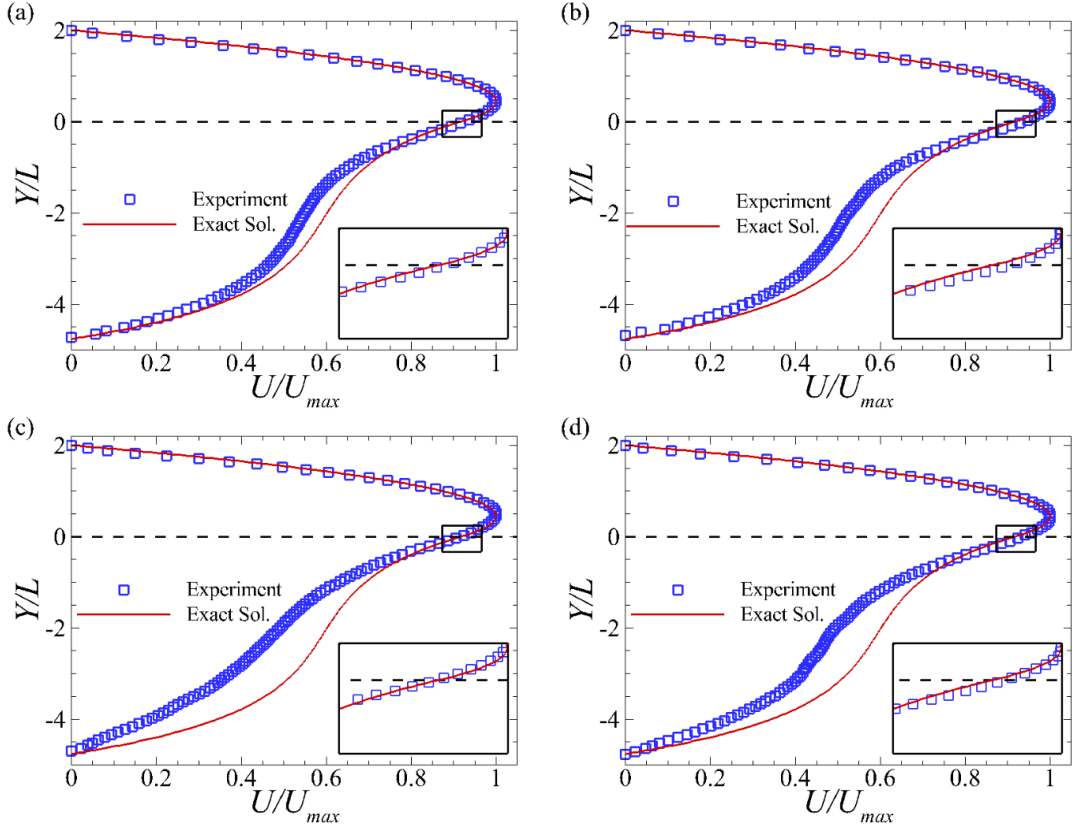

FIGURE 7. The area averaged, normalized velocity profiles for (a) pure suspending fluid, suspensions of (b) $\phi_b = 0.01$, (c) $\phi_b = 0.03$, and (d) $\phi_b = 0.05$. The red line indicates the velocity profile obtained from the exact solution of pure Newtonian fluid moving over porous surfaces. The black dashed line is the interface between the porous media and the free-flow region.

The velocity at the interface averaged in time and space gives the apparent slip velocity over the rods, $U_s = U_{ave}|_{y/L=0}$ [67]. The slip velocity was normalized using two techniques, one of which compares the slip velocity to global parameters such as the maximum velocity, $U_{max}$. Fig. 8(b) shows these values derived from the velocity profiles in the $P_1$ plane, $P_2$ plane, and the full averaged plane for a solvent and dilute suspensions. As can be seen, by increasing particle concentrations the normalized slip velocity increases for the plane within the porous media and it and it decays for the plane on top of the porous media. The pure fluid is the outlier in this data set for the two data collection planes by having a 7.2% decrease in the $P_1$ plane and a 12.3% increase in the $P_2$ plane. However, by averaging the data taken on $P_1$ and $P_2$ planes the variation of the dimensionless slip velocity for increasing the concentration of particles is trivial. The black line shows the value from the exact solutions of the Newtonian fluid. This shows a close agreement to the values obtained from the experimental averaged profile.

The second method to normalize the slip velocity is to use local parameters such as the shear rate, $\dot{\gamma}$, and the square root of the permeability, $\sqrt{K}$. This expression is known as the dimensionless slip parameter, $U_s/\dot{\gamma}\sqrt{K}$, and is considered to be more useful in describing the flow physics inside the porous media due to the fact that it depends on the local conditions as opposed to the far-field velocity profile [11, 24]. This dimensionless slip parameter is shown in Fig. 8(c) for the values from the velocity profile in the $P_1$ plane, $P_2$ plane and the full averaged profile. The dimensionless slip parameter shows opposite trends to that of the dimensionless slip velocity in the $P_1$ and $P_2$ planes. As the concentration increases the dimensionless slip parameter decreases in the $P_1$ planes, and it increases in the $P_2$ plane. The full averaged profile shows similar results for the solvent and the dilute suspensions. The value obtained from the exact solution is close to the experimental values; however, it is interesting to note that, for this parameter, the exact solution is closer to the values for the 3% and 5% suspensions.

For Newtonian fluid over porous media models with height ratios of $\delta = 0.78$ and $\delta = 2.55$, Agelinchaab et al. (2006) found that the dimensionless slip velocity of approximately 1 and 2, respectively. They also reported that these dimensionless values were not dependent on the porosity or the space between the rods. The maximum values of the dimensionless slip velocity for pure fluid flow in an open channel over a glass fiber porous material has been found to be $U_s/\dot{\gamma}\sqrt{K} = 2.3 - 14.3$ [9]. The height ratio in our experiments is $\delta = 4.63$, which then reflects the higher dimensionless slip velocity (i.e., 3.84 for the solvent). This data revealed that very dilute suspensions passing over and through a porous media model still have similar behavior to that of pure Newtonian fluid. Moreover, as can be seen in Fig. 8(c), the plane within the rods shows a large deviation in the dimensionless slip velocity, but it has a more linear trend for the plane on top of the rods. This could be due to the fact that the shear rate varies within the rods as has been observed previously for laminar flow of pure fluid over the riblets [68].

We have already shown that the presence of porous media in leu of a solid boundary has produced a nonzero velocity (slip velocity) at the interface between the free flow region and the porous media. This phenomenon also produces a distance from where the velocity would be zero when extrapolating the interfacial velocity gradient of the overlying fluid, known as the slip length, $l_{slip}$. Mathematically, this can be defined as,

$$l_{slip} = \frac{U_s}{dU_{ave}|_{y/L=0}} \qquad (4.4)$$

These values for the solvent and the dilute suspensions are shown in Figure 8(d) for the $P_1$, $P_2$ and full averaged profiles. The trends for the $P_1$ and $P_2$ planes are similar to those seen in the dimensionless slip parameter due to the increase in concentration. The slip length for the solvent and very dilute suspensions show close agreement with the value derived from the analytic solution, shown by the black line on the figure.

The parameters obtained by the analysis of a very dilute suspension over and through a porous media model is of great interest both theoretically and practically. Practical examples of these applications include; oil recovery and manufacturing processes of advanced composites. In these examples, the focus of examination needs to be on the flow near the permeable surface and the extent of the impacts of the boundary perturbations inside the porous media. Therefore, the obtained experimental data allows us to also provide useful data on the phenomenological constants used in the existing models for slow steady-state pure fluid over and within a porous medium (i.e., Brinkman's equation) and explore their validity for very dilute suspensions. To continue examining the behavior of dilute suspensions over and inside a porous media model, we defined the parameters including the average Darcy velocity, $U_D$, the slip coefficient, $\alpha$, and the penetration depth, $\delta$.

To model the boundary condition at the interface between the porous media and a pure fluid region Beavers and Joseph (1967) proposed eq. (1.1). In order to calculate the slip coefficient, $\alpha$, using this equation one needs to first determine the Darcy velocity, $U_D$. This is challenging because the experimental profiles do not show a clear plug flow inside the porous media that is typically used to determine the average Darcy velocity [8]. To solve this conundrum, we employed two different methods to determine $U_D$.

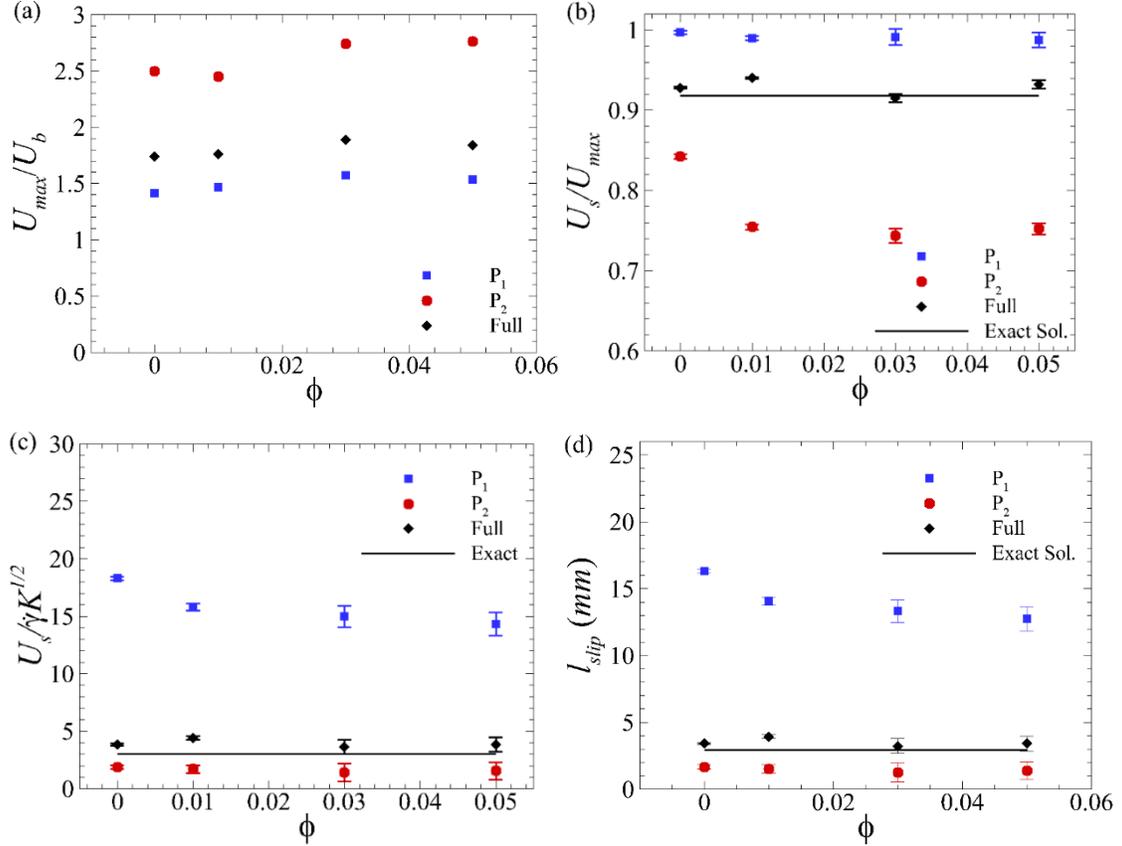

FIGURE 8. (a) The maximum velocity normalized by the bulk velocity. The slip velocity normalized by (b) the maximum area-averaged velocity (with a maximum error of 1%), and (c) the shear rate and $\sqrt{K}$, for the planes $P_1$ and $P_2$ as well as the averaged profile (with a maximum error of 7%). (d) The slip length for the two data collection planes and the averaged profile for various flow suspensions, with a maximum error of 7.1%. The black line on figures (b-d) indicates the results obtained using the exact solution for the pure fluid over porous media. The error bars show the uncertainty derived based on Coleman and Steel (1995) and Sciacchitano and Wieneke (2016).

The first technique was to find $U_D$ through the use of Darcy's law that is dependent on the pressure drop, $dp/dx$ [10, 69, 70],

$$U_D = \frac{-K}{\mu}\frac{dp}{dx}. \tag{4.5}$$

To evaluate $U_D$ using this equation, we first needed to define the pressure drop above the porous media. We derived a relationship for the pressure drop for the solvent using the simplified 1D Navier-Stokes equation for pure Newtonian fluid within parallel plates given by [52],

$$\mu\frac{d^2u}{dy^2} - \frac{dp}{dx} = 0. \tag{4.6}$$

Here, $u$ is the streamwise component of velocity which depends only on the $y$ coordinate, $\mu$ is the viscosity for the solvent fluid, and $dp/dx$ indicates the pressure drop. By integrating eq. (4.6), the velocity profile can be given by

$$u(y) = \frac{1}{\mu}\frac{dp}{dx}\left(\frac{y^2}{2} - Ly\right) + U_s\left(1 - \frac{y}{2L}\right). \tag{4.7}$$

To incorporate the presence of the porous media, the lower boundary condition required to solve the integrate is $u(y = 0) = U_s$. To isolate the pressure drop term, eq. (4.7) was integrated across the free flow region to obtain the flow rate, $Q$. Rearranging this relationship provided pressure drop as a function of the flow rate, $Q$, the slip velocity, $U_s$, the half gap of the free flow region, $L$, and the fluid viscosity, $\mu$.

$$\frac{dp}{dx} = \frac{3\mu(Q - U_s L)}{2L^3}. \tag{4.8}$$

We were able to use experimental values for $Q$ and $U_s$ to calculate the pressure gradient using eq. (4.8) and subsequently solving for the Darcy velocity using eq. (4.5). We define the phrase, "solvent exact equations", to describe obtaining the Darcy velocity using eq. (4.8) and eq. (4.5). These equations are only valid for pure Newtonian fluid; however, we have used the solvent exact equations to calculate the Darcy velocity for the dilute suspensions, but we replaced the solvent viscosity with the Krieger viscosity [55].

Another method to obtain the Darcy velocity for 2D porous media is by using the velocity profile. Previous literature has used an average of the velocity within the porous media to define the Darcy velocity [71]. It was determined that this region of the velocity profile which is averaged together depends on the porosity of the porous media. We were able to compare the averaged velocity within the porous media across varying $Y/L$ locations to the Darcy velocity obtained through the solvent exact equations for the pure fluid case. From this we could define a set region within the porous media (i.e., $-2.82 < Y/L < -1.67$) that we could average across to get the Darcy velocity which would match the result from the solvent exact solution. Since this region is dependent on porosity, the region was constant for the other dilute suspensions.

The values for $U_D$ from both techniques can be seen in Fig. 9(a). The results from the two methods are similar for the solvent and for the 1% suspension, in fact, for 1% suspension the values between the velocity profiles and the exact solvent solutions are the same. However, they deviate as the concentration of suspension increases (i.e., for 3% and 5% suspensions). Examining the suspensions, the experiments show a decrease in $U_D$ where the solvent exact equations show a slight increase in this velocity. This might be due to the impact of particle-rod interaction; thus, reducing the velocity within the porous media model that is not reflected in the pure fluid calculations.

Using the $U_D$ from the two methods and the shear rate $(dU/dy|_{y=0^+})$ calculated at the interface, the slip coefficient $\alpha$ can be calculated from eq. (1.1) for all test cases. The slip coefficients calculated using the Darcy velocity obtained through averaging of the velocity profile and the solvent exact solutions can be seen in Fig. 9(b). The results from the solvent exact equations show a similar trend to that of the results from the experimental profile. For the pure fluid and the 1% suspension experiments the slip coefficient derived from the two methods are almost identical, however, at the higher concentration these two values differ. It is clear from this figure, that at the higher concentrations the calculated values using the solvent exact equations cannot capture the proper flow physics of a suspension within a porous media and more detailed modeling will need to be developed for this.

Finally, examined the penetration depth $\delta$, defined as the transitional region between the free-flow and the Darcy velocity region within the porous media [8]. We considered $\delta$, as the distance from the interface to the point on the velocity profile where the velocity decays to $1.01 U_D$ [21, 72]. The penetration depth of the Darcy velocity obtained by the averaging of part of the velocity profile for all test cases can be seen in Fig. 9(c). These values are found to be very similar for the solvent and the dilute suspensions, with an average value for all cases of $2.33 \pm 0.043$ mm. This led to the conclusion that the penetration depth is more dependent on the properties of the porous media than the concentration of the suspension. The error for this data, since it is a location-based variable, is 0.035, i.e., the half of the interrogation size [71].

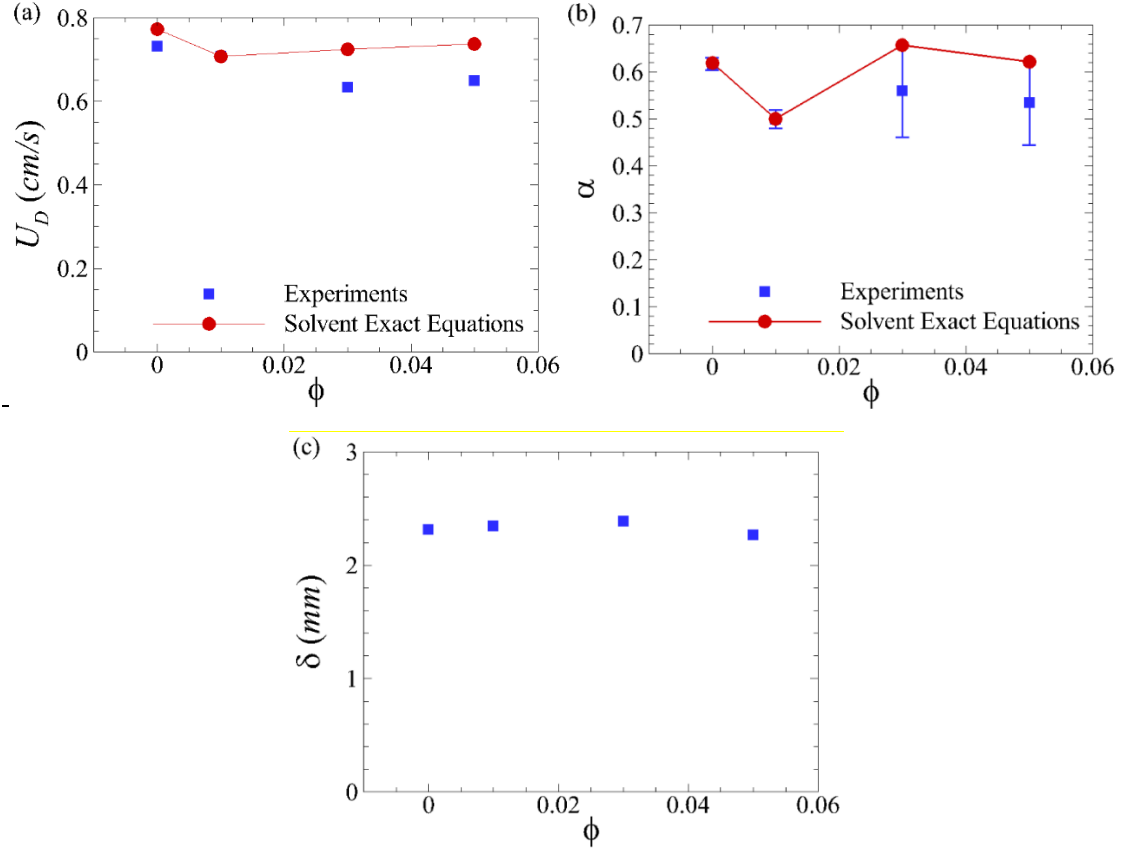

FIGURE 9. The (a) Darcy velocity, (b) slip coefficient, and (c) penetration depth for the solvent, 1%, 3%, and 5% concentrated suspensions. The red lines on the Darcy velocity (a) and slip coefficient (b) figures symbolize the results from the solvent exact equations (i.e., eq. (4.5)) and (4.8)). Average error for the Darcy velocity is 4% as calculated through the methods of Coleman and Steel (1995). The error for the penetration depth is half the interrogation resulting in 1.5% error. These resulting errors were too small to view on the figures.

It should be noted that the slip coefficient, $\alpha$, introduced by Beavers and Joseph, has been the subject of investigation in multiple literatures [8, 10, 73, 74]. For instance, $\alpha$ was shown to yield an identical value of the slip velocity at the fluid-porous interface where for an open flow can be defined as $\alpha = \sqrt{\mu/\mu_e}$ where $\mu$ and $\mu_e$ are the fluid and the effective viscosity, respectively [72, 75]. The effective viscosity was characterized as the viscosity of the fluid moving through the porous media. The shear flow was studied over square and hexagonal porous media at different orientations to the flow direction. It was observed that $\mu > \mu_e$ and $\mu < \mu_e$ when the flow is perpendicular and parallel to the flow, respectively [76, 77]. Later on, a relationship between the viscosity and effective viscosity based on the porosity of the porous media has been defined as $\mu_e/\mu = 1/\varepsilon$ [78]. Using PIV techniques for Newtonian fluid over a porous model (e.g., brush-like structures) inside a Couette device, found that $\mu = \mu_e$ [6]. In some cases where the porosity is sufficiently low the flow within the porous media (the Darcy velocity) is very low and can be neglected in eq. (1.1), this yields a relationship of $\alpha = \left(U_s/\dot{\gamma}\sqrt{K}\right)^{-1}$ [11, 25]. Statistical approaches were derived for the case where the porous media consists of packed spheres [69, 79]. [69] Saffman (1971) determined that for densely packed spheres/ suspensions the effective viscosity is $\mu_e/\mu = 1/(1 - 2.5\phi)$. In the current study, we used our experimental data to obtain slip coefficients for suspending fluid and various suspensions.

Previous research involving porous media has studied the impact of the porous media on the flow rate within the channel [10, 52, 78, 80]. This analysis attempted to quantify how various porous media models would affect the flow structure. Beavers and Joseph proposed an increase ratio of the flow rates due to the presence of porous media. The so-called fractional increase ratio of the mass flow rates used the Darcy's equation and can be defined as,

$$\Phi = \frac{Q_p - Q_i}{Q_i} = \frac{3(\sigma + \alpha)}{\sigma(1 + 2\alpha\sigma)} \tag{4.9}$$

where $\alpha$ is the slip coefficient and $\sigma$ is the dimensionless permeability parameter [52]. This fractional increase in mass flow rate was studied extensively for pure Newtonian fluid over porous media using the exact solution and a direct numerical simulations (DNSs) [52]. They compared the results for $\Phi$ obtained using the exact solutions, DNS results, Beavers and Joseph (BJ) (1967) [10] experiments, BJ analytical analysis, and Brinkman equation for cases with comparable $\alpha$ and $\varepsilon$ values. It was shown that all of these approaches provide similar results for a various range of $\sigma$ [52].

Using $\alpha$ values presented in Fig. 9(b) and the dimensionless permeability parameter $\sigma = L/\sqrt{K}$, we were able to characterize the fractional increase of mass flow rate for the solvent and the dilute suspensions as shown in Fig. 10(a). In this graph, the black line shows the fractional increase in the mass flow rate using the solvent exact solution (i.e., eq. (4.5) and (4.8)). Although this equation was developed for pure Newtonian fluid, it presents a good agreement to the results obtained from very dilute suspensions. It should be noted that the error bars are quite large for these data points, which is due to the differentiation and multiple manipulations of the error derived from the velocity measurements.

To gain further insight into how the presence of the particles affect the flow within the channel containing the porous media, the flow rate in the free-flow region, $Q$, was calculated for all experiments. The ratio of the flow rate in the suspension-flow region to the solvent-flow region is plotted in Fig. 10(b). One might conclude that by adding more particles to the flow and in the existence of a porous model, the flow rate inside the channel decays. Further analysis of the impact of particles over and through the rods in the porous media model is the subject of our further investigations.

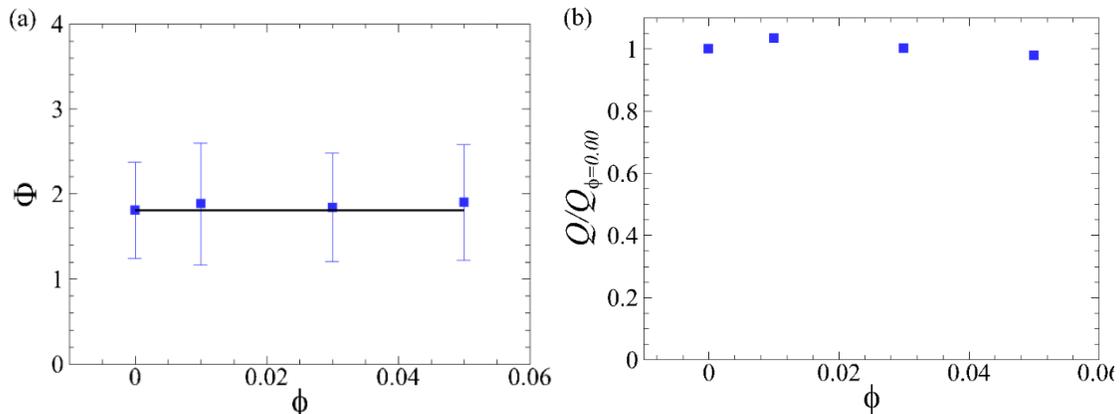

FIGURE 10. (a) The fractional increase of the mass flow rate due to the presence of a porous media compared to the increase in concentration [10]. The error bars are calculated through error propagation analysis of the equations [59]. (b) The ratio of the free flow region flow rate of the dilute suspensions over the pure fluid.

## 5. Conclusions

We have presented the first detailed analysis of pressure-driven flow of very dilute suspensions of rigid, spherical, non-Brownian, and non-colloidal suspensions over a three-dimensional porous structure using the PIV and refractive index matching techniques. We have considered the suspending fluid, 1%, 3%, and 5% concentrated suspensions and porous medium consisting of regular solid structures of a general shape in our analysis. The porous model consists of rods with a height of 5mm, a porosity of 0.9, and a permeability of $7.93 \times 10^{-7} m^2$. The main contribution of the current paper is to provide the foundation and the tools to understand suspensions over a 3D porous media model. In particular, we neglected the impact of inertia and considered only creeping flow in our experiments which allowed us to compare them with extensive works that have been conducted on pure Newtonian fluid flowing over porous media as well as suspension flows over smooth surfaces in creeping flows. These experiments were compared to the exact solution velocity profile obtained analytically using the coupling of the continuity and Navier-Stokes equation in the free flow region and VANS equations inside the porous media. The obtained exact solution velocity profile showed good agreement with the experimental results, especially in the free-flow region. Within the porous media, however, as the concentration increased the velocity profiles began to depart from the exact solutions. We believe this is due to the impact of the rods-particles, particle-particle interactions inside the porous media. Nonetheless, the present experimental data could be used as a benchmark to refine the modeling of suspensions flowing over porous structures and the related boundary conditions at the suspension-porous interface. This study explores in detail the velocity variations of dilute suspensions over and within a porous media. Analysis of the concentration profiles and the particle migration over the porous media is the subject of our current investigation.

Based on the presented results the research outcomes are as follows:

1. For the suspensions considered in this study, the velocity profiles within the rods are not uniform. This effect can be seen more in the higher concentrations. This is because of the particle-rod effects that are similar to the data taken for pure fluid motion over porous models reported in previous literatures by [24, 25, 64]. Moreover, for the thickness and the porosity of the porous media model considered in this work, the boundary layers produced at the interface and on the lower surfaces of the channel might extend to results in non-uniform velocity profiles. These impacts have been proven to vanish for pure fluid motion over porous media by decreasing porosity of the porous media [12, 25]. It is likely that these boundary layer effects would considerably vary in the flow of concentrated suspensions.

2. The ratio of slip velocity to the maximum area-averaged velocity is relatively constant for the solvent, 1%, 3%, and 5% concentrated suspensions. This shows that the velocity above the porous media is similar for pure fluid and various suspensions. The dimensionless slip parameter, $U_s/\dot{\gamma}\sqrt{K}$, and the slip length, $l_{slip}$, for the experimental results are relatively invariant to change in concentration. This was because the velocity profiles for the solvent and dilute suspensions are all still similar at the interface; therefore, the properties at this interface should be comparable across all cases tested.

3. Although the comparison of the velocity profiles obtained experimentally and from the exact solutions provide rather good agreement, specifically in the free-flow region, there is still a discrepancy between the experiments and the exact solutions within the porous media. In fact, the particles in the experiments migrating through the porous media model play a critical role, which results in varying the velocity profiles. The experimental results show that the velocity profiles are strongly dependent on the suspension concentration flowing through the porous media, but it is relatively independent from the concentration in the free-flow region. Nonetheless, this analysis is a proof of concept that one might characterize the velocity profile of 1% suspensions flowing over and inside a porous medium model containing of rods using the exact solutions of the pure Newtonian flow over and through porous layer when the porous medium property, the thickness of the layer, the particle size, shape and concentration as well as the geometry of the channel are known.

4. The slip coefficient and dimensionless slip parameter remain relatively unchanged by the increase in concentrations. However, the slip coefficient is lower in the dilute suspensions than the solvent. Since these values are still comparable it further supports the idea that for very dilute suspensions, they have similar behavior to that of the pure Newtonian fluid.

5. Further examination of the experimental data, provided useful information on the phenomenological constants used in the existing models for pure Newtonian fluid and explore their validity for very dilute suspensions. Our analysis shows that the penetration depth $\delta$, and the fractional increase of the mass flow rate due to the presence of a porous media $\Phi$, are almost the same for pure solvent fluid and the suspensions.

**Acknowledgements**

This work has been supported partially by National Science Foundation Award No. 1854376 and partially by Army Research Office Award No. W911NF-18-1-0356. We also thank Dr. Changwoo Kang for valuable discussion in terms of the exact solutions of the pure fluid.